\documentclass[onecolumn,amsmath,amssymb,showkeys]{revtex4}

\usepackage{amsmath}
\usepackage{graphicx}
\usepackage{graphics}
\usepackage{dcolumn}
\usepackage{bm}

\def\BEq{\begin{equation}}
\def\EEq{\end{equation}}
\def\BEqA{\begin{eqnarray}}
\def\EEqA{\end{eqnarray}}
\def\BEn{\begin{enumerate}}
\def\EEn{\end{enumerate}}
\def\BWT{\begin{widetext}}
\def\EWT{\end{widetext}}

\def\a{\alpha}

\begin{document}


\title{Anisotropic Keldysh interaction}

\author{Andrei Galiautdinov}
 \affiliation{
Department of Physics and Astronomy, 
University of Georgia, Athens, Georgia 30602, USA}
\email{ag@physast.uga.edu}

\date{\today}

\begin{abstract}

We generalize the classic calculations by Rytova and Keldysh of screened Coulomb 
interaction in semiconductor thin films to systems with anisotropic permittivity tensor. 
Explicit asymptotic expressions for electrostatic potential energy of interaction in 
the weakly anisotropic case are found in closed analytical form. The case of strong 
in-plane anisotropy, however, requires evaluation of the inverse Fourier transform of
$1/(k+Ak_x^2+Bk_y^2)$, which, at present, remains unresolved.

\end{abstract}


\maketitle


\section{Introduction}

The important role played by the dielectric screening in determining excitonic properties 
of various two-dimensional semiconductor heterostructures has been the subject of
numerous investigations over the last several decades (for recent studies see, for example, 
\cite{cudazzo2010strong,cudazzo2011dielectric, chernikov2014exciton,low2014plasmons, 
wang2015highly,chaves2015anisotropic,latini2015excitons,pedersen2016exciton,
trolle2017model,hichri2017dielectric,szyniszewski2017binding,mostaani2017diffusion,
cavalcante2018electrostatics}). 
A particularly interesting direction of current experimental research involves 
perovskite chalcogenide films whose in-plane dielectric 
anisotropy gives rise to some rather unusual optical behavior
\cite{niu2018giant,niu2018mid}. 
Past theoretical work on two-dimensional dielectric screening involved
various {\it ab initio} calculations \cite{cudazzo2010strong,cudazzo2011dielectric},  
the use of the nonlinear Thomas-Fermi model \cite{low2014plasmons}, 
\cite{latini2015excitons}, the modified Mott-Wannier approach \cite{pedersen2016exciton}, 
the transfer matrix method \cite{cavalcante2018electrostatics},
as well as various approaches based on effective mass approximation 
\cite{chaves2015anisotropic,hichri2017dielectric},
Here we pursue what is likely the simplest possibility --- generalization to anisotropic 
films of classic calculations by Rytova \cite{Rytova1967} and Keldysh 
\cite{keldysh1979coulomb, keldysh1997excitons}. The motivation for this approach is rather 
obvious: we want to get a better sense of how the famous {\it isotropic} form of screened
electrostatic interaction energy,
\BEq 
V(\rho)=(\pi q q'/\epsilon d) [H_0(\rho/\rho_0)-Y_0(\rho/\rho_0)],
\EEq 
is modified under the minimal number of microscopic assumptions.
In what follows, we provide the general expression for the Fourier image of 
the anisotropic potential in momentum space, and analytically 
work out in real space the {\it weakly} anisotropic case only. Interested readers 
are invited to improve on that calculation by exploring the {\it strongly} anisotropic 
scenario. 

\section{General considerations}

The electrostatic potential energy of interaction between charges $q$ and $q'$ 
located at $(\bm{\rho},z)$ and $({\bf 0}, z')$ ($z>z'$, $\bm{\rho}=(x,y)$)
inside an {\it anisotropic} semiconductor film of thickness $d$ surrounded
by two isotropic media with dielectric constants $\epsilon_1$ and $\epsilon_2$
is given by (see Appendix for derivation and Fig.\ \ref{fig:6}; 
compare with \cite{Rytova1967, keldysh1979coulomb})
\BEq
\label{eq:aRK3D}
V(\bm{\rho},z,z')=\int \frac{d^2{\bf k}}{(2\pi)^2} e^{i{\bf k}\cdot{\bm{\rho}}}
V({\bm{k}},z,z'),
\EEq
with
\begin{align}
V({\bm{k}},z,z')
=
\frac{4{\pi} qq'}{\epsilon_{z}}
\frac{
\cosh \left[{\tilde k}\left(\frac{d}{2}-z\right)+{\tilde \eta}_2\right]
\cosh \left[{\tilde k}\left(\frac{d}{2}+z'\right)+{\tilde \eta}_1\right]
}{
{\tilde k}\sinh [{\tilde k}d+{\tilde \eta}_1+{\tilde \eta}_2]
},
\end{align}
where
\begin{align}
\label{eq:eta1,2}
{\tilde \eta}_{1,2}
=
\frac{1}{2}\ln
\left(\frac{\epsilon_{z}\tilde{k}+\epsilon_{1,2} k}
{\epsilon_{z}\tilde{k}-\epsilon_{1,2} k}\right), 
\quad
{\tilde k}=\sqrt{\frac{\epsilon_{x}k_x^2+\epsilon_{y}k_y^2}{\epsilon_{z}}},
\quad
k=\sqrt{k_x^2+k_y^2},
\end{align}
and the axes of the coordinate system $(x,y,z)$ coincide with the principal axes 
of the film's permittivity tensor, 
$\epsilon = {\rm diag}(\epsilon_{x},  \epsilon_{y}, \epsilon_{z})$. 
In the most interesting for practical applications scenario, 
$\epsilon_{1,2}\ll \epsilon_{x},  \epsilon_{y}, \epsilon_{z}$, and, 
for distances $\rho \gg |z-z'| \sim d$, the main contribution to the integral in 
(\ref{eq:aRK3D}) comes from ${\bf k}$ satisfying $kd\ll k\rho \lesssim 1$. 
Under these conditions, 
${\tilde k}d\ll1$, ${\tilde \eta}_{1,2} \approx \epsilon_{1,2}k/(\epsilon_{z}{\tilde k})$, 
and, with the dependence on $z$ and $z'$ disappearing, we get the two-dimensional 
form of the interaction,
\begin{align}
\label{eq:aRK2D}
V(\bm{\rho})
&=
\frac{4\pi qq'}{(2{\pi})^2 d}
\int 
\frac{d^2{\bf k} \,
e^{i{\bf k}\cdot{\bm{\rho}}}
}{
\epsilon_{x}k_x^2+\epsilon_{y}k_y^2
+(\epsilon_{1}+\epsilon_{2})\frac{k}{d}
}.
\end{align}

At this point it is convenient to introduce two ``screening'' lengths,
\begin{align}
\label{eq:screeningLengths}
\rho_{0x}\equiv \frac{\epsilon_x d}{\epsilon_{1}+\epsilon_{2}},
\quad
\rho_{0y}\equiv \frac{\epsilon_y d}{\epsilon_{1}+\epsilon_{2}},
\end{align}
characterizing polarizability of the film in the $x$ and $y$ directions,
respectively, and write the interaction (\ref{eq:aRK2D}) in the form
\begin{align}
\label{eq:aRK2DscreeningLengths}
V(\bm{\rho})
&=
\frac{qq'}{{\pi} (\epsilon_{1}+\epsilon_{2}) }
\int 
\frac{d^2{\bf k} \,
e^{i{\bf k}\cdot{\bm{\rho}}}
}{k\, \epsilon({\bf k})
},
\end{align}
where $\epsilon({\bf k})$ is the dielectric function, formally defined by
\BEq
\epsilon({\bf k}) 
= 1+\frac{1}{k}
\left(\rho_{0x} k_x^2+{\rho_{0y} k_y^2}\right),
\EEq
which generalizes the standard isotropic result \cite{cudazzo2011dielectric}.
In Ref.\ \cite{berkelbach2013theory}, for the case of surrounding vacuum in 
the {\it isotropic} scenario, the authors have numerically verified that the 
screening length of a monolayer can be calculated with good accuracy 
on the basis of Eq.\ (\ref{eq:screeningLengths})
provided the dielectric contrast is large and the relevant dielectric constant 
of the monolayer is the in-plane component of the permittivity 
tensor of the bulk material. We take that as an indication that the Keldysh model 
is a good approximation to realistic experimental situations and hypothesize 
that its anisotropic generalization proposed here should work reasonably well 
even for samples of monolayer thickness.

The problem thus reduces to the calculation of a two-dimensional Fourier integral,
\BEq
F(x,y)
=
\int 
\frac{dk_x dk_y}{(2\pi)^2}
\frac{
e^{i(k_x x+k_y y)} 
}{
k+Ak_x^2+Bk_y^2
}, 
\EEq
with $A,B >0$. 
To that end, working in polar coordinates, we write,
\begin{align}
\label{eq:VpolarCoords1}
V(\bm{\rho})&=
\frac{qq'}{{\pi}(\epsilon_{1}+\epsilon_{2})}
\int_0^{\infty}  \int_0^{2\pi} 
\frac{ dt d\theta \, e^{it\cos\theta} }
{t[\rho_{0x}\cos^2 (\theta+\alpha)+\rho_{0y}\sin^2 (\theta+\alpha)]+\rho}
\nonumber \\
&=
\frac{2qq'}{(\epsilon_{1}+\epsilon_{2})\rho_{0}}
\int_0^{\infty} 
dt\,
\frac{I(t,\a,a,b)}{t +b},
\end{align}
where $t=k\rho$, $\alpha$ is the angle between the position vector 
${\bm{\rho}}$ and the positive $x$-axis, as shown in Fig.\ \ref{fig:6},
\BEq
\label{eq:myFunction}
I(t,\a,a,b)=\frac{1}{2{\pi}}\int_0^{2\pi}d\theta
\frac{
e^{it\cos\theta} 
}{
1
- 
\varepsilon^2(t,a,b)
\cos^2 (\theta+\alpha)
},
\EEq
and
\BEq
\varepsilon^2(t,a,b)
\equiv  
\frac{a t}{t +b},
\quad
a\equiv 1-\frac{\rho_{0x}}{\rho_{0y}},
\quad
b \equiv \frac{\rho}{\rho_{0}},
\quad
{\rho}_{0}\equiv \rho_{0y},
\EEq
with $a\in (-\infty,1]$ playing the role of the anisotropy parameter; the greater 
the $|a|$, the greater the anisotropy, with $a=0$ corresponding to the isotropic 
case. 

Without loss of generality, we may assume that
$0 < \rho_{0x} \leq \rho_{0y}$, and thus $0\leq a <1$.
Then, since $t\geq 0$, we have $0\leq \varepsilon^2(t) <1$.
Taking into account the well-known Fourier series expansion 
\cite{mikhlin1964integral},
\begin{align}
\frac{1}{1-\varepsilon^2 \cos^2 \chi}
=
\frac{1}{\sqrt{1-\varepsilon^2}}
\bigg[
1
+
2\sum^{\infty}_{n=1}
\bigg(\frac{\varepsilon}{1+\sqrt{1-\varepsilon^2}}\bigg)^{2n}\cos(2n\chi)
\bigg],
\end{align}
and using the fact that
$\int_0^{2\pi}d\theta \, e^{it\cos\theta} \sin(2n\theta)=0$,
we get for the $\theta$-integral in (\ref{eq:myFunction}) the 
asymptotic multipole series,
\begin{align}
\label{eq:I(t)}
I(t,\a,a,b)
&=
\frac{1}{\sqrt{1-\varepsilon^2(t)}}
\frac{1}{2{\pi}}\int_0^{2\pi}
d\theta \,
e^{it\cos\theta}
\nonumber \\
&\quad
+ 
\frac{2}{\sqrt{1-\varepsilon^2(t)}}\sum^{\infty}_{n=1}
\bigg(\frac{\varepsilon(t)}{1+\sqrt{1-\varepsilon^2(t)}}\bigg)^{2n}
\bigg(
\frac{1}{2{\pi}}\int_0^{2\pi}
d\theta \,
e^{it\cos\theta} \cos(2n\theta)
\bigg)\cos(2n\alpha)
\nonumber \\
&=
 \frac{ J_0(t)}{\sqrt{1-\varepsilon^2(t)}}
\nonumber \\
&\quad
+
\frac{2}{\sqrt{1-\varepsilon^2(t)}}
\bigg(\frac{\varepsilon(t)}{1+\sqrt{1-\varepsilon^2(t)}}\bigg)^{2}
\, \frac{tJ_0(t)-2 J_1(t)}{t}
\, \cos (2 \alpha )
\nonumber \\
&\quad
+
\frac{2}{\sqrt{1-\varepsilon^2(t)}}
\bigg(\frac{\varepsilon(t)}{1+\sqrt{1-\varepsilon^2(t)}}\bigg)^{4}
\, \frac{t \left(t^2-24\right) J_0(t)-8 \left(t^2-6\right) J_1(t)}{t^3} 
\, \cos(4\alpha)
\nonumber \\
&\quad
+
\frac{2}{\sqrt{1-\varepsilon^2(t)}}
\bigg(\frac{\varepsilon(t)}{1+\sqrt{1-\varepsilon^2(t)}}\bigg)^{6}
\,
\frac{t \left(t^4-144 t^2+1920\right) J_0(t)-6 \left(3 t^4-128 t^2+640\right) 
J_1(t)}{t^5}
\, \cos(6\alpha) 
+ \, \dots,
\nonumber \\
\end{align}
where $J_0$ and $J_1$ are the Bessel functions of the first kind.

\section{Weak anisotropy}

\label{sec:weakAnisotropy}

In a rather straightforward manner (for a better approach see 
Sec.\ \ref{sec:renormalizedKeldysh}), assuming $0\leq a\ll 1$ and 
treating $\varepsilon^2(t,a,b)$ as a small parameter 
in (\ref{eq:I(t)}), we get, in lowest order,
\begin{align}
\label{eq:I(t)-weak}
I(t)
\approx
  J_0(t)
+ \frac{\varepsilon^2(t)}{2} J_0
+ \frac{\varepsilon^2(t)}{2}  \bigg(J_0(t)-\frac{2J_1(t)}{t}\bigg)\, 
\cos(2\alpha),
\end{align}
and, thus,
\begin{align}
\label{eq:aRK2D-weak}
V(b,\a)&=
\frac{2qq'}{(\epsilon_{1}+\epsilon_{2})\rho_{0}}
\int_0^{\infty} 
dt
\bigg\{
\frac{J_0(t)}{t+b}
+ 
\frac{a}{2} 
\bigg[
\frac{tJ_0(t)}{(t +b)^2} 
+ 
\frac{tJ_0(t)-2J_1(t)}{(t +b)^2} 
\, \cos(2\alpha)
\bigg]
\bigg\}.
\end{align}
\begin{figure}[!ht]
\includegraphics[angle=0,width=1\linewidth]{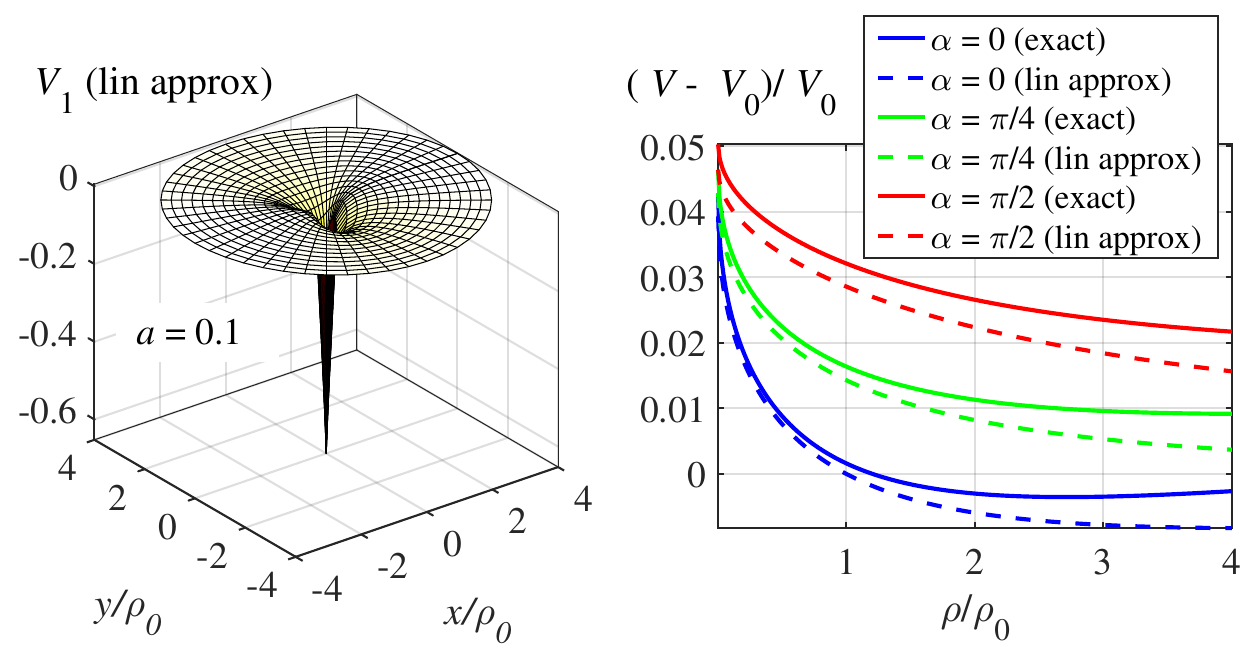}
\caption{ \label{fig:1}  
Graphs of $V_1$ in units of 
${|qq'|}/(\epsilon_{1}+\epsilon_{2})\rho_{0}$,
and $V_1/V_0$, with $a=0.1$, in the weakly anisotropic case, calculated 
on the basis of Eqs.\ (\ref{eq:aRK2Dweak}) and (\ref{eq:linApproxKeldysh})
and direct numerical integration in Eq.\ (\ref{eq:VpolarCoords1}).
}
\end{figure}
Performing the remaining $t$-integration we find, 
\BEq
\label{eq:linApprox}
V(b,\alpha)= V_0(b) + V_1(b,\alpha), 
\EEq
where
\begin{align}
\label{eq:linApproxKeldysh}
V_0(b,\alpha) 
= \frac{{\pi}qq'}{(\epsilon_{1}+\epsilon_{2})\rho_{0}}[{H}_0(b)-Y_0(b)]
\end{align}
is the standard Keldysh-Rytova result, and
\begin{align}
\label{eq:aRK2Dweak}
V_1(b,\alpha)
&=
\frac{a}{2}\frac{\pi qq'}{(\epsilon_{1}+\epsilon_{2})\rho_{0}}
\left[{H}_0(b) - Y_0(b) + \frac{2}{\pi} b + b Y_1(b) - b {H}_1(b) \right]
\nonumber \\
& \quad
+
\frac{a}{2}\frac{\pi qq'}{(\epsilon_{1}+\epsilon_{2})\rho_{0}}
\left[
\frac{2}{\pi} b^3
+ b \left[\left(b^2-2\right) Y_1(b)+ b Y_0(b)\right]
-  b \left[\left(b^2-2\right) {H}_1(b)+b {H}_0(b)\right]
-\frac{4}{\pi}
\right]\frac{\cos (2 \alpha ) }{b^2}
\end{align}
is the linear correction whose graph is shown in Fig.\ \ref{fig:1}
(assuming $qq'<0$). In the above, various $H_i$ and $Y_i$ 
denote the Struve and Neumann functions, respectively. 

For $b\ll1$, or ${d}\ll{\rho}\ll \rho_{0}$, 
we get
\begin{align}
\label{eq:aRK2Dweak1}
V_0 
&= 
\frac{2qq'}{(\epsilon_{1}+\epsilon_{2})\rho_{0}} 
\left[\ln \left(\frac{2}{b}\right)-\gamma \right]
=
\frac{2qq'}{ (\epsilon_{1}+\epsilon_{2})\rho_{0}}
\left[
\ln\left(\frac{2\rho_{0}}{\rho}\right)
-\gamma
\right],
\\
V_1
&=
\frac{aqq'}{ (\epsilon_{1}+\epsilon_{2})\rho_{0}}
  \left[\ln \left(\frac{2}{b}\right)-\gamma -1-\frac{\cos (2 \alpha )}{2}\right]
=
\frac{aqq'}{(\epsilon_{1}+\epsilon_{2})\rho_{0}} 
\left[
\ln\left(\frac{2\rho_{0}}{\rho}\right)
-\gamma-1
-\frac{\cos(2 \alpha)}{2} \right],
\end{align}
where $\gamma \approx 0.577216$ is the Euler constant. 
Since $\int_0^{2\pi}\cos (2 \alpha )d\alpha =0$, the excitonic ground state 
energy in this case experiences a first order shift,
\BEq
\Delta E_0 = \frac{2\pi aqq'}{(\epsilon_{1}+\epsilon_{2})\rho_{0}}  
\int_0^{\infty} \left(\ln \left(\frac{2}{b}\right)-\gamma -1\right) 
|\psi_0(b)|^2 \, b\, db,
\EEq
where $\psi_0(b)$ is the unperturbed axially symmetric ground state 
wave function. 
On the other hand, for $b\gg1$, or 
${\rho}\gg \rho_{0}$, 
Eqs.\ (\ref{eq:linApproxKeldysh}) and (\ref{eq:aRK2Dweak}) reproduce the 
standard Coulomb asymptotics,
\BEq
V(\rho)
= \frac{2 qq'}{(\epsilon_{1}+\epsilon_2)\rho_0}
\left( \frac{1}{b} - \frac{a\cos (2\alpha)}{b^2}\right)
\rightarrow \frac{2 qq'}{(\epsilon_{1}+\epsilon_2)\rho}.
\EEq

\begin{figure}[!ht]
\includegraphics[angle=0,width=1\linewidth]{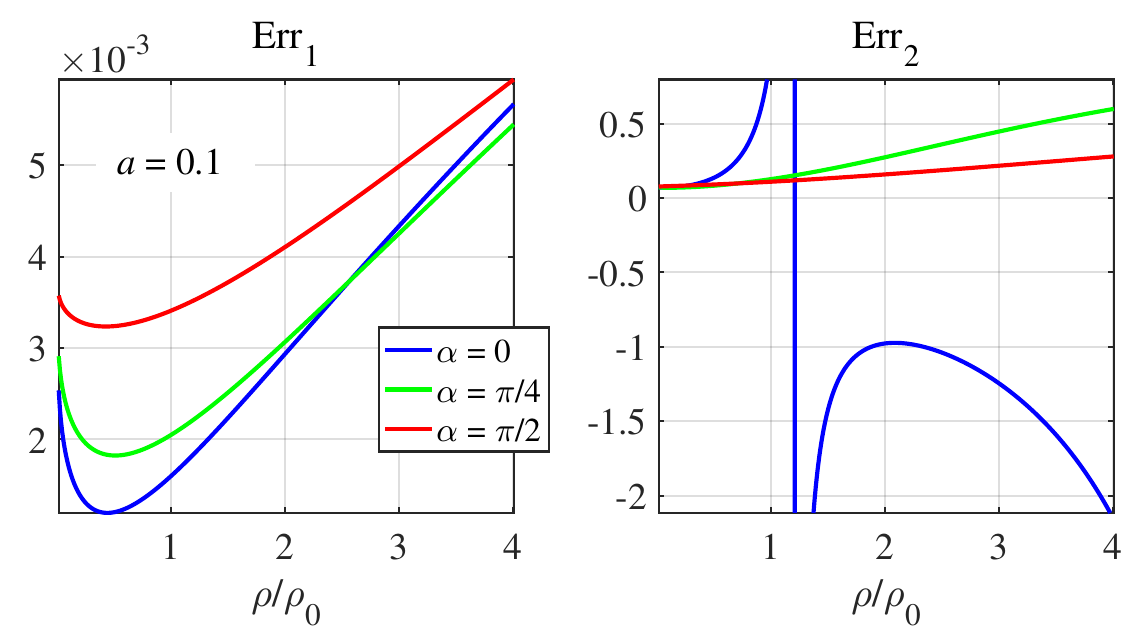}
\caption{ \label{fig:2}  
Graphs of the relative errors defined in Eqs.\ (\ref{eq:error1})
and (\ref{eq:error2}). The vertical asymptote (in blue) is at the
point with $\alpha = 0$, for which $V=V_0$ (exact Keldysh value); 
compare with Fig.\ \ref{fig:1}.
}
\end{figure}
To get a sense of the error involved in this linear approximation, we define two
relative errors by 
\BEq
\label{eq:error1}
{\rm Err}_1\equiv \frac{V_{\rm exact} - V_{\rm approx}}{V_{\rm exact}}
= \frac{V_{1\,{\rm exact}} - V_1}{V_{\rm exact}}
\EEq
and
\BEq
\label{eq:error2}
{\rm Err}_2\equiv \frac{V_{\rm exact} - V_{\rm approx}}{V_{\rm exact}-V_0}
= \frac{V_{1\,{\rm exact}} - V_1}{V_{1\,{\rm exact}}},
\EEq
respectively, with $V_{1\,{\rm exact}}\equiv V_{\rm exact}-V_0$.
Here, $V_{\rm approx}\equiv V_0 + V_1$ is calculated on the basis of 
Eqs.\ (\ref{eq:linApproxKeldysh}) and (\ref{eq:aRK2Dweak}), 
and $V_{\rm exact}$ is found by direct numerical integration of the double integral
in (\ref{eq:VpolarCoords1}). The corresponding results are summarized in 
Fig.\ \ref{fig:2}. Notice that for all $\rho$ the error ${\rm Err}_1$ is greatest 
for points with $\a=\pi/2$. The error ${\rm Err}_2$ is particularly troublesome,
as the blue curve clearly indicates.

\section{Weak anisotropy: Renormalized Keldysh interaction}
\label{sec:renormalizedKeldysh}

\begin{figure}[!ht]
\includegraphics[angle=0,width=1\linewidth]{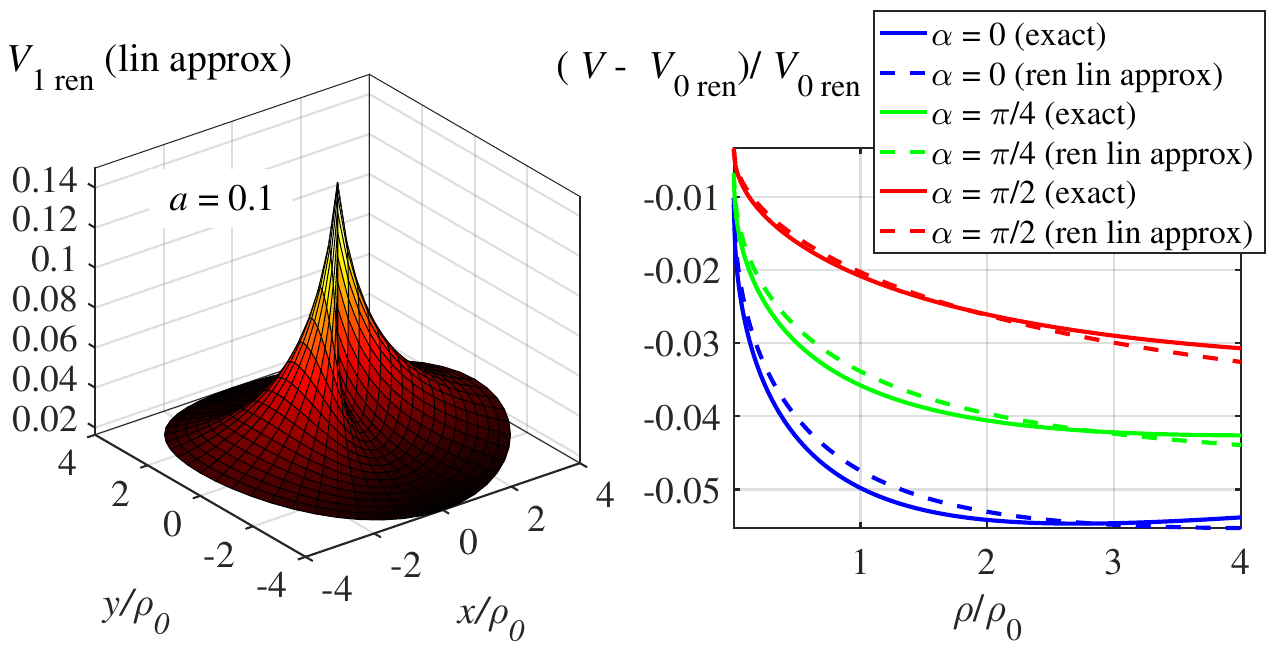}
\caption{ \label{fig:3}  
Graphs of $V_{1\, {\rm ren}}$ in units of 
${|qq'|}/(\epsilon_{1}+\epsilon_{2})\rho_{0}$,
and $V_{1\, {\rm ren}}/V_{0\, {\rm ren}}$, with $a=0.1$, 
in the weakly anisotropic case, calculated 
on the basis of Eqs.\ (\ref{eq:V0ren}) and (\ref{eq:V1ren})
and direct numerical integration in Eq.\ (\ref{eq:VpolarCoords1}).
Compare with Fig.\ \ref{fig:1}.
}
\end{figure}

A better linear approximation can be achieved by ``renormalizing'' the zeroth 
order Keldysh contribution, $V_0$, as follows: we re-write the monopole term in 
(\ref{eq:I(t)}) as shown below,
\begin{align}
\label{eq:I(t)renormalized}
I
&=
  \frac{ J_0(t)}{\sqrt{1-a}} 
+ 
\underbrace{
\bigg[
\bigg(
\frac{ J_0(t)}{\sqrt{1-\varepsilon^2(t)}} -  \frac{ J_0(t)}{\sqrt{1-a}}
\bigg)
+
\frac{2}{\sqrt{1-\varepsilon^2(t)}}
\bigg(\frac{\varepsilon(t)}{1+\sqrt{1-\varepsilon^2(t)}}\bigg)^{2}
\, \frac{tJ_0(t)-2 J_1(t)}{t}
\, \cos (2 \alpha )
+ \, \dots
\bigg]
}_{\sim {\cal O}(a)},
\end{align}
and expand everything in square brackets to linear (leading!) order in $a$.
The potential then becomes
\begin{align}
V(b,\a)
&=
\frac{2qq'}{(\epsilon_{1}+\epsilon_{2})\rho_{0}}
\int_0^{\infty} 
dt\,
\bigg\{
\frac{1}{\sqrt{1-a}}\frac{J_0(t)}{t+b} 
+ \frac{a}{2}\, 
\bigg[
- \frac{b J_0(t)}{(t+b)^2}
+ \frac{ tJ_0(t)-2 J_1(t)}{(t+b)^2}
\, \cos (2 \alpha )
\bigg]
\bigg\},
\end{align}
which should be compared with (\ref{eq:aRK2D-weak}). We then find
\begin{align}
\label{eq:VrenormalizedLinear}
V(b,\a)
&=V_{0\, {\rm ren}}+V_{1\, {\rm ren}},
\end{align}
where
\begin{align}
\label{eq:V0ren}
V_{0\, {\rm ren}}(b,\alpha)&=
 \frac{{\pi}qq'}{(\epsilon_{1}+\epsilon_{2})\rho_{0}}
\frac{{H}_0(b)-Y_0(b)}{\sqrt{1-a}}
\end{align}
is the {\it renormalized} isotropic Keldysh term, and
\begin{align}
\label{eq:V1ren}
V_{1\, {\rm ren}}(b,\alpha)&=
\frac{a}{2}\frac{\pi qq'}{(\epsilon_{1}+\epsilon_{2})\rho_{0}}
\left[\frac{2}{\pi} b + b Y_1(b) - b {H}_1(b) \right]
\nonumber \\
& \quad
+
\frac{a}{2}\frac{\pi qq'}{(\epsilon_{1}+\epsilon_{2})\rho_{0}}
\left[
\frac{2}{\pi} b^3
+ b \left[\left(b^2-2\right) Y_1(b)+ b Y_0(b)\right]
-  b \left[\left(b^2-2\right) {H}_1(b)+b {H}_0(b)\right]
-\frac{4}{\pi}
\right]\frac{\cos (2 \alpha ) }{b^2}
\end{align}
is the corresponding linear correction consisting of a linear monopole and a linear
dipole contributions, Fig.\ \ref{fig:3}. 
Now in the $b\ll1$ limit we get 
\begin{align}
V_{0\, {\rm ren}}
&= 
\frac{2qq'}{(\epsilon_{1}+\epsilon_{2})\rho_{0}} 
\frac{\ln \left(\frac{2}{b}\right)-\gamma }{\sqrt{1-a}}
\end{align}
and a perfectly reasonable first order correction
\begin{align}
V_{1\, {\rm ren}}
&=
-\frac{aqq'}{ (\epsilon_{1}+\epsilon_{2})\rho_{0}}
  \left[1+\frac{\cos (2 \alpha )}{2}\right],
\end{align}
which does not contain the logarithmic term. The excitonic ground state 
energy in this case undergoes a simple first order shift,
\BEq
\Delta E_{0\, {\rm ren}} = -\frac{aqq'}{ (\epsilon_{1}+\epsilon_{2})\rho_{0}}.
\EEq

Notice that our renormalization procedure eliminates logarithmic terms in all orders 
of the {\it monopole} perturbation, not just the first one.
For example, keeping the second order monopole contribution in square brackets in
Eq.\ (\ref{eq:I(t)renormalized}) would add the term
\begin{align}
V_{2\, {\rm ren}}^{({\rm monopole})}
&=\frac{3a^2}{16} 
\frac{ \pi qq'}{ (\epsilon_{1}+\epsilon_{2})\rho_{0}}
 \left[\frac{8}{\pi}b +  b^2 Y_0(b)-  b^2 {H}_0(b) +3 b  Y_1(b)-3 b {H}_1(b)\right]
\end{align}
to the potential in (\ref{eq:VrenormalizedLinear}), which in the $b\ll1$ limit is just
\begin{align}
V_{2\, {\rm ren}}^{({\rm monopole})}
&=-\frac{9a^2}{8} \frac{qq'}{ (\epsilon_{1}+\epsilon_{2})\rho_{0}}.
\end{align}

\begin{figure}
\includegraphics[angle=0,width=1\linewidth]{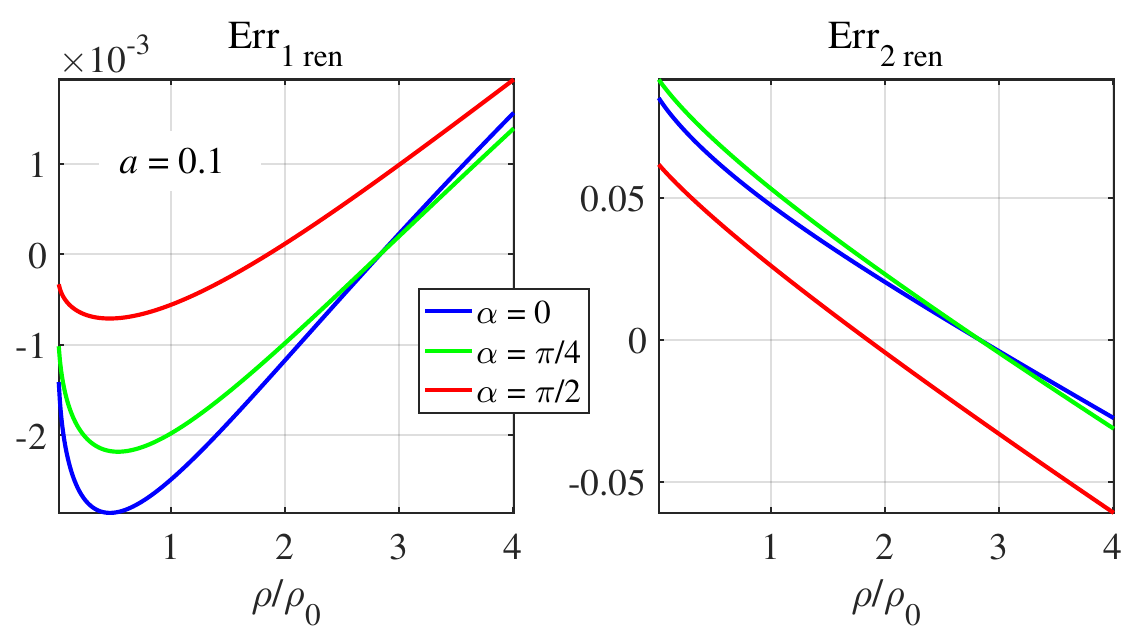}
\caption{ \label{fig:4}  
Graphs of the relative errors defined in Eqs.\ (\ref{eq:error1ren})
and (\ref{eq:error2ren}). Compare with Fig.\ \ref{fig:2}.
}
\end{figure}
\begin{figure}[!ht]
\includegraphics[angle=0,width=1\linewidth]{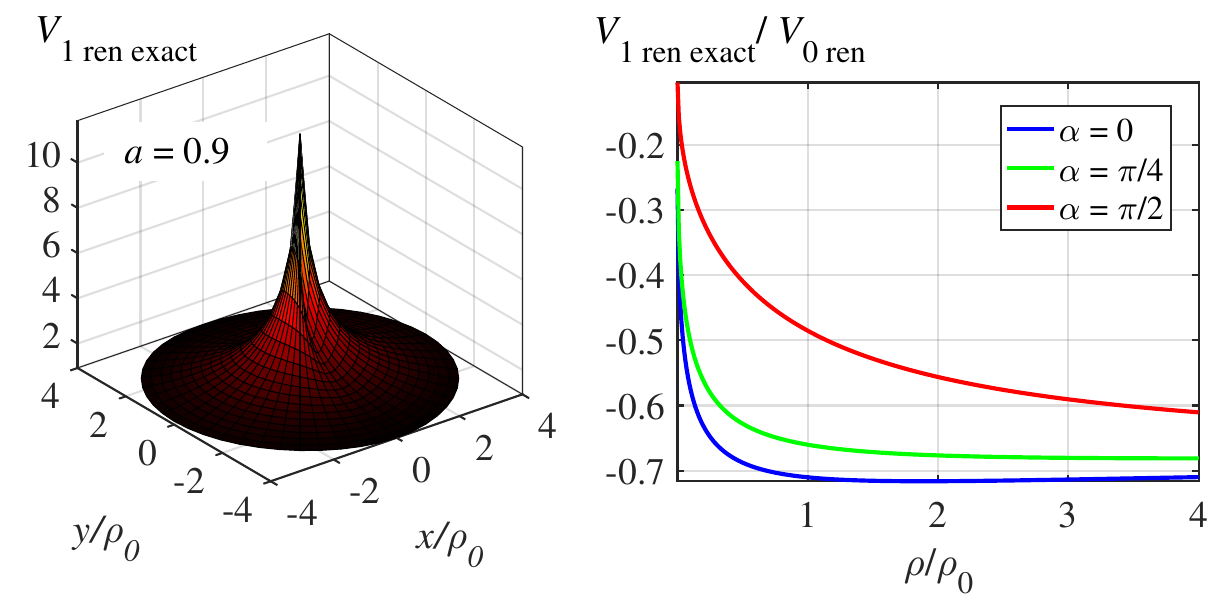}
\caption{ \label{fig:5}  
Graphs of $V_{1\, {\rm ren}\,{\rm exact}}$ in units of 
${|qq'|}/(\epsilon_{1}+\epsilon_{2})\rho_{0}$
and $V_{1\, {\rm ren}\,{\rm exact}}/V_{0\, {\rm ren}}$, with $a=0.9$, in 
the strongly anisotropic case (calculated numerically 
on the basis of Eqs.\ (\ref{eq:VpolarCoords1}) 
and (\ref{eq:V0ren})).
}
\end{figure}
Returning to the linear approximation (\ref{eq:VrenormalizedLinear}), we again define two 
relative errors,
\BEq
\label{eq:error1ren}
{\rm Err}_{1\, {\rm ren}}\equiv \frac{V_{\rm exact} - V_{\rm ren \, approx}}{V_{\rm exact}}
= \frac{V_{1\, {\rm ren}\,{\rm exact}} - V_{1\, {\rm ren}}}{V_{\rm exact}},
\EEq
and
\BEq
\label{eq:error2ren}
{\rm Err}_{2\, {\rm ren}}
\equiv \frac{V_{\rm exact} - V_{\rm ren \, approx}}{V_{\rm exact}-V_{0\, {\rm ren}}}
= \frac{V_{1\, {\rm ren}\,{\rm exact}} - V_{1\, {\rm ren}}}{V_{1\, {\rm ren}\,{\rm exact}}},
\EEq
with $V_{1\, {\rm ren}\,{\rm exact}}\equiv V_{\rm exact}-V_{0\, {\rm ren}}$ and
$V_{\rm ren \, approx}\equiv V_{0\, {\rm ren}} + V_{1\, {\rm ren}}$. 
The corresponding numerical results presented in Fig.\ \ref{fig:4} show
that our revised approximation scheme is indeed superior to the one used in 
Sec.\ \ref{sec:weakAnisotropy}.

Finally, we also performed numerical simulations in the extreme anisotropic regime,
as shown in Fig.\ \ref{fig:5}. In this case, the ``correction'' 
$V_{1\,{\rm ren}\,{\rm exact}}$
becomes comparable to $V_{0\,{\rm ren}}$, and the linear approximation breaks down.

\section{Summary}
The classic Keldysh-Rytova formula for screened Coulomb interaction in semiconductor
thin films has been generalized by taking into account the anisotropy of the layer's
dielectric permittivity tensor. The Fourier image of the anisotropic potential in
momentum space, as well as the linear correction to the isotropic potential in 
real space, have been worked out in closed analytical form. The case of strong 
in-plane anisotropy, however, remains unresolved due to the appearance 
of the function $I(t,\a,a,b)$ (see Eqs.\ (\ref{eq:myFunction}) and (\ref{eq:I(t)})),
whose explicit analytical expression is not known.  

\begin{acknowledgments}

The author thanks Robert Zaballa for useful discussions.

\end{acknowledgments}

\section*{APPENDIX: Momentum space representation}
\label{sec:Appendix}

\begin{figure}
\includegraphics[angle=0,width=1\linewidth]{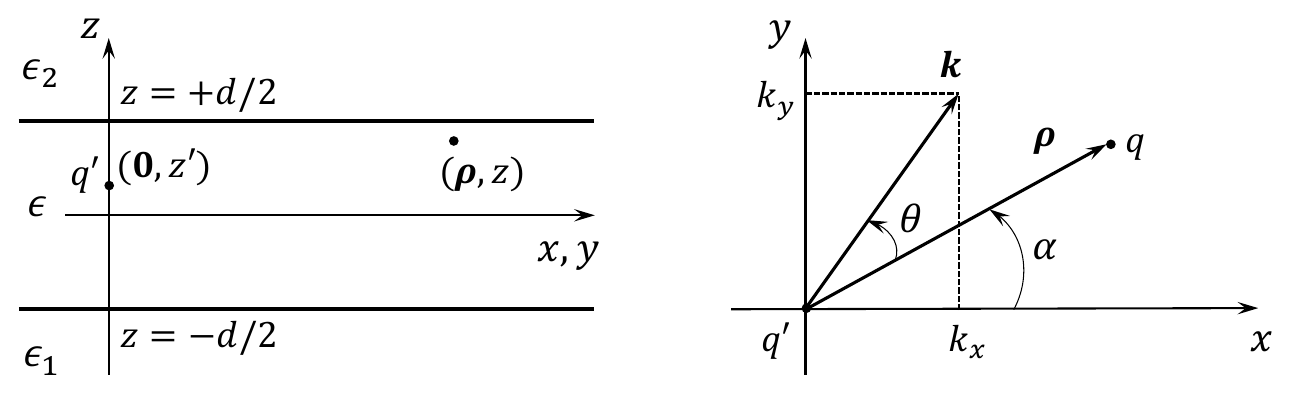}
\caption{ \label{fig:6}  
Left: Semiconductor film geometry; the $(x,y,z)$ axes 
coincide with the principal axes of the dielectric permittivity tensor of the film, 
$\epsilon$.
Right: Mutual orientation of vectors ${\bf k}$ and ${\bm{\rho}}$ used
in Eq.\ (\ref{eq:aRK2D}). 
}
\end{figure}

Following \cite{Rytova1967} and \cite{keldysh1979coulomb}, 
we consider a geometry in which
the anisotropic semiconductor film occupies the region of space $-d/2\leq z \leq d/2$, 
as shown in Fig.\ \ref{fig:6}. The half-space $z<-d/2$ (the substrate) is filled 
with an isotropic medium whose dielectric constant is $\epsilon_1$,
while the half-space $z>d/2$ with an isotropic medium whose dielectric 
constant is $\epsilon_2$. 

We are assuming that the axes of the coordinate system $(x,y,z)$ 
coincide with the principal axes of the film's dielectric permittivity tensor. 
The electrostatic potential at point ${\bf r}\equiv (\bm{\rho},z)=(x,y,z)$ 
due to charge $q'$ located at ${\bf r}'=(0,0,z')$ satisfies in regions 1, 2, 
and 3 (the film) the following system of equations:
\begin{align}
\label{eq:1}
\left(\frac{\partial^2 }{\partial x^2}
+\frac{\partial^2 }{\partial y^2}
+\frac{\partial^2 }{\partial z^2}\right)\phi_1({\bf r},{\bf r}')&=0,
\\
\label{eq:2}
\left(\frac{\partial^2 }{\partial x^2}
+\frac{\partial^2 }{\partial y^2}
+\frac{\partial^2 }{\partial z^2}\right)\phi_2({\bf r},{\bf r}')&=0,
\\
\label{eq:3}
\left(\epsilon_{x}\frac{\partial^2 }{\partial x^2}
+\epsilon_{y}\frac{\partial^2 }{\partial y^2}
+\epsilon_{z}\frac{\partial^2 }{\partial z^2}\right)\phi_3({\bf r},{\bf r}')
&=-4{\pi}q' \delta({\bf r}-{\bf r}'),
\end{align} 
with the boundary conditions at the interfaces,
\begin{align}
\label{eq:bc3}
&(\phi_2-\phi_3)_{z=\frac{d}{2}}=0,
\quad
\left(\epsilon_2\frac{\partial \phi_2}{\partial z}
-\epsilon_{z}\frac{\partial \phi_3}{\partial z}\right)_{z=\frac{d}{2}}=0,
\\
\label{eq:bc1}
&(\phi_3-\phi_1)_{z=-\frac{d}{2}}=0,
\quad
\left(\epsilon_{z}\frac{\partial \phi_3}{\partial z}
-\epsilon_{1}\frac{\partial \phi_1}{\partial z}\right)_{z=-\frac{d}{2}}=0,
\end{align}
and the boundary conditions at the two infinities,
\BEq
\label{eq:bc0}
\phi_2\vert_{z\rightarrow+\infty}=0,
\quad
\phi_1\vert_{z\rightarrow-\infty}=0.
\EEq

Fourier transforming,
\begin{align}
\label{eq:ft321}
\phi_{1,2,3}({\bf r},{\bf r}')
&=\int \frac{d^2{\bf k}}{(2\pi)^2}
e^{i{\bf k}\cdot{\bm{\rho}}}\phi_{1,2,3}({\bf k},z,z'),
\quad
{\bf k}=(k_x, k_y),
\end{align}
and substituting into (\ref{eq:1}), (\ref{eq:2}), and (\ref{eq:3}), we get 
the following equations for the corresponding Fourier components,
\begin{align}
\label{eq:k1}
\left(\frac{\partial^2 }{\partial z^2}-k^2\right)\phi_1(k,z,z')&=0,
\\
\label{eq:k2}
\left(\frac{\partial^2 }{\partial z^2}-k^2\right)\phi_2(k,z,z')&=0,
\\
\label{eq:k3}
\left(\frac{\partial^2 }{\partial z^2}-{\tilde k}^2\right)\phi_3({\tilde k},z,z')
&=-\frac{4{\pi}q'}{\epsilon_{z}} \delta(z-z'),
\end{align}
where
\BEq
k\equiv |{\bf k}|=\sqrt{k_x^2+k_y^2}, 
\quad
{\tilde k}\equiv \sqrt{\frac{\epsilon_{x}k_x^2+\epsilon_{y}k_y^2}{\epsilon_{z}}}.
\EEq
Conditions at infinity, (\ref{eq:bc0}), combined with Eqs.\ (\ref{eq:k1}) 
and (\ref{eq:k2}) give
\begin{align}
\label{eq:phik2}
\phi_{2}(k)&=A_2(z') e^{-kz},
\quad
\frac{\partial \phi_{2}(k)}{\partial z}=-kA_2(z') e^{-kz},
\\
\label{eq:phik1}
\phi_{1}(k)&=A_1(z') e^{kz},
\quad
\frac{\partial \phi_{1}(k)}{\partial z}=kA_1(z') e^{kz},
\end{align}
while Eq.\ (\ref{eq:k3}) gives, for $z\neq z'$,
\begin{align}
\label{eq:phik3neq}
\phi_{3}({\tilde k}, z\neq z')&=A_3(z') e^{{\tilde k}z}+B_3(z') e^{-{\tilde k}z},
\end{align}
and, at $z=z'$, the jump discontinuity in the $z$-derivative,
\BEq
\label{eq:phik3'jump}
\frac{\partial \phi_{3}({\tilde k})}{\partial z}\bigg|^{z=z'+0}_{z=z'-0}
=-\frac{4{\pi} q'}{\epsilon_{z}}.
\EEq
Imposing the boundary conditions at the $z=+d/2$ interface,
\begin{align}
\label{eq:bck2}
A_2(z')e^{-{kd}/{2}}
&=
A_3(z') e^{{{\tilde k}d}/{2}}+B_3(z') e^{-{{\tilde k}d}/{2}},
\\
\label{eq:bck2'}
-\epsilon_2 kA_2(z')e^{-{kd}/{2}}
&=
\epsilon_{z} {\tilde k} 
\left[A_3(z') e^{{{\tilde k}d}/{2}}-B_3(z') e^{-{{\tilde k}d}/{2}}\right],
\end{align}
we get
\begin{align}
\label{eq:phik3>}
\phi_{3}({\tilde k})\vert_{z> z'}
&=
A_3(z') 
\left[
e^{{\tilde k}z}+e^{2{\tilde \eta}_2+{\tilde k}(d-z)}
\right],
\end{align}
where
\BEq
\label{eq:eta2}
{\tilde \eta}_2
\equiv
\frac{1}{2}\ln
\left(\frac{\epsilon_{z}\tilde{k}+\epsilon_2 k}{\epsilon_{z}\tilde{k}-\epsilon_2 k}\right).
\EEq
Similarly, imposing the boundary conditions at the $z=-d/2$ interface, we get
\begin{align}
\label{eq:bck1}
{\tilde A}_3(z') e^{-{{\tilde k}d}/{2}}+{\tilde B}_3(z') e^{{{\tilde k}d}/{2}}
&=
A_1(z')e^{-{kd}/{2}},
\\
\label{eq:bck1'}
\epsilon_{z} {\tilde k} 
\left[{\tilde A}_3(z') e^{-{{\tilde k}d}/{2}}-{\tilde B}_3(z') e^{{{\tilde k}d}/{2}}\right]
&=
\epsilon_1 kA_1(z')e^{-{kd}/{2}},
\end{align}
and, after defining
\BEq
\label{eq:eta1}
{\tilde \eta}_1
\equiv
\frac{1}{2}\ln
\left(\frac{\epsilon_{z}\tilde{k}+\epsilon_1 k}{\epsilon_{z}\tilde{k}-\epsilon_1 k}\right),
\EEq
find
\begin{align}
\label{eq:phik3<}
\phi_{3}({\tilde k})\vert_{z<z'}
&=
{\tilde A}_3(z') 
\left[
e^{{\tilde k}z}+e^{-2{\tilde \eta}_1-{\tilde k}(d+z)}
\right].
\end{align}
Now, for $z=z'$, Eqs.\ (\ref{eq:phik3'jump}), (\ref{eq:phik3>}), and
(\ref{eq:phik3<}) give
\begin{align}
A_3(z') 
\left[
e^{{\tilde k}z'}+e^{2{\tilde \eta}_2+{\tilde k}(d-z')}
\right]
-
{\tilde A}_3(z') 
\left[
e^{{\tilde k}z'}+e^{-2{\tilde \eta}_1-{\tilde k}(d+z')}
\right]&=0,
\\
{\tilde k}A_3(z') 
\left[
e^{{\tilde k}z'}-e^{2{\tilde \eta}_2+{\tilde k}(d-z')}
\right]
-
{\tilde k}{\tilde A}_3(z') 
\left[
e^{{\tilde k}z'}-e^{-2{\tilde \eta}_1-{\tilde k}(d+z')}
\right]
&=-\frac{4{\pi}q'}{\epsilon_{z}},
\end{align}
resulting in
\begin{align}
{\tilde A}_3(z') &= A_3(z') 
\frac{
e^{{\tilde k}z'}+e^{2{\tilde \eta}_2+{\tilde k}(d-z')}
}{
e^{{\tilde k}z'}+e^{-2{\tilde \eta}_1-{\tilde k}(d+z')}
},
\end{align}
and
\begin{align}
A_3(z') 
&=
\frac{4{\pi}q'}{\epsilon_{z}{\tilde k}}
\frac{
e^{-{\tilde k}d/2-{\tilde \eta}_2}
\cosh \left[{\tilde k}(d/2+z')+{\tilde \eta}_1\right]
}
{
2
\sinh\left[{\tilde k}d+{\tilde \eta}_1+{\tilde \eta}_2\right]
}.
\end{align}
Taking into account that
\begin{align}
e^{{\tilde k}z}+e^{2{\tilde \eta}_2+{\tilde k}(d-z)}
&=
2e^{{\tilde k}d/2+{\tilde \eta}_2}\cosh\left[{\tilde k}(d/2-z)+{\tilde \eta}_2\right],
\end{align}
we get
\begin{align}
\phi_3({\tilde k},z>z')
=
\frac{4{\pi} q'}{\epsilon_{z}}
\frac{
\cosh \left[{\tilde k}\left(\frac{d}{2}-z\right)+{\tilde \eta}_2\right]
\cosh \left[{\tilde k}\left(\frac{d}{2}+z'\right)+{\tilde \eta}_1\right]
}{
{\tilde k}\sinh [{\tilde k}d+{\tilde \eta}_1+{\tilde \eta}_2]
}.
\end{align}
For $z<z'$, a similar calculation results in
\begin{align}
\phi_3({\tilde k},z<z')
=
\frac{4{\pi} q'}{\epsilon_{z}}
\frac{
\cosh \left[{\tilde k}\left(\frac{d}{2}-z'\right)+{\tilde \eta}_2\right]
\cosh \left[{\tilde k}\left(\frac{d}{2}+z\right)+{\tilde \eta}_1\right]
}{
{\tilde k}\sinh [{\tilde k}d+{\tilde \eta}_1+{\tilde \eta}_2]
}.
\end{align}

\end{document}